\documentclass[twocolumn,floatfix,groupedaddress]{revtex4}

\usepackage{graphicx}
\usepackage{amsmath}
\usepackage{epsfig}
\usepackage{txfonts}

\newcommand{\be}{\begin{equation}}
\newcommand{\e}{\end{equation}}
\newcommand{\beml}{\begin{subequations}}
\newcommand{\eml}{\end{subequations}}
\newcommand{\beq}{\begin{eqnarray}}
\newcommand{\eq}{\end{eqnarray}}
\newcommand{\ba}{\begin{array}}
\newcommand{\ea}{\end{array}}

\begin{document}

\title{Spin qubits in graphene quantum dots}

\author{Bj{\"o}rn Trauzettel, Denis V. Bulaev, Daniel Loss, and Guido Burkard}

\affiliation{Department of Physics and Astronomy, University of Basel,
Klingelbergstrasse 82, CH-4056 Basel, Switzerland}

\date{January 2007}

\maketitle

{\bf The electron spin is a very promising candidate for a 
solid-state qubit \cite{loss98}.
Major experimental breakthroughs have been achieved in recent years
using quantum dots formed in semiconductor heterostructures based on
GaAs technology \cite{elze04,hans05,pett05,kopp06}. 
In such devices, the major sources of spin decoherence
have been identified as the spin-orbit interaction, coupling the spin to
lattice vibrations \cite{khae00,khae01,golo04},
and the hyperfine interaction of the electron spin
with the surrounding nuclear spins 
\cite{burk99,erli01,khae02,cois04,john05,kopp05}.
%
Therefore, it is
desirable to form qubits in quantum dots based on other materials,
where spin-orbit coupling and hyperfine interaction are considerably weaker \cite{min06}. 
It is well known
that carbon-based materials such as  nanotubes or graphene are excellent
candidates.  This is so because spin-orbit coupling is weak in carbon
due to its relatively low atomic weight, and because natural carbon consists
predominantly of the zero-spin isotope $^{12}$C, for which 
the hyperfine interaction is absent.  
Here we show how to form spin qubits in graphene.
A crucial requirement to achieve this goal is to find quantum dot states
where the usual valley degeneracy is lifted. We show that this problem
can be avoided in quantum dots with so-called armchair boundaries.
We furthermore show that spin qubits in graphene can not
only be coupled (via Heisenberg exchange) between nearest neighbor 
quantum dots but also
over  long distances. This remarkable feature is a direct consequence of the
Klein paradox being a distinct property of the quasi-relativistic spectrum of
graphene.
Therefore, the proposed system is ideal for fault-tolerant 
quantum computation, and thus for scalability, 
since it offers a low error rate due to weak decoherence,
in combination with a high error threshold due to the possibility of 
long-range coupling.}

\begin{figure}
\begin{center}
\includegraphics[scale=0.25]{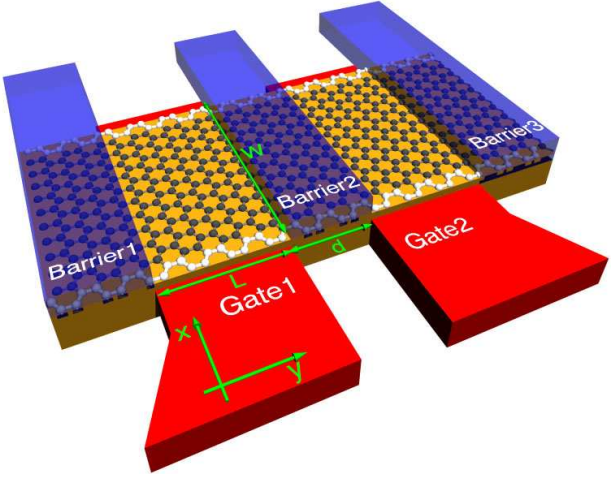} 
\caption{\label{setup2} 
{\bf Schematic of a graphene double quantum dot.}
Each dot is assumed to have length $L$ and width $W$. 
The structure is based on a ribbon of graphene (grey) with 
semiconducting armchair edges (white). 
Confinement is achieved by tuning the voltages applied to the ``barrier'' 
gates (blue) to appropriate values such that bound states exist. 
Additional gates (red) allow to shift the energy levels of the dots.
Virtual hopping of electrons through barrier 2 (thickness $d$) gives rise to
a tunable exchange coupling $J$ between two electron 
spins localized in the left and the right dot.
The exchange coupling is then used to generate universal two-qubit gates.}
\end{center}
\end{figure}

Only very recently, the fabrication of a single layer of graphene and the 
measurement of its electric transport properties have been achieved 
\cite{novo04,novo05,zhan05}.
Two fundamental problems need to be overcome before graphene can be used to
form spin qubits and to operate one or two of them as proposed in 
Refs.~\cite{loss98,burk99}: (i) It is
difficult to create a tunable quantum dot in graphene because of the absence
of a gap in the spectrum. The phenomenon of Klein tunnelling makes it hard to
confine particles \cite{chei06,domb99,kats06}. 
(ii) Due to the valley degeneracy that exists in
graphene \cite{mccl56,seme84,divi84}, 
it is non-trivial to form two-qubit gates using Heisenberg exchange
coupling for spins in tunnel-coupled dots. Several attempts have been made to
solve the problem (i) such as to use suitable transverse states in graphene
ribbons to confine electrons \cite{silv06}, to combine single and bilayer
regions of graphene \cite{nils06}, or to achieve confinement by using
inhomogeneous magnetic fields \cite{dema06}. The problem (ii) has not been
recognized up to now. Here we propose a setup which solves both
problems (i) and (ii) at once. Similar to Ref.~\cite{silv06} we choose to
confine electrons by using suitable transverse states in a ribbon of graphene,
{\it cf.} Fig.~\ref{setup2}. In particular, we assume {\it semiconducting
  armchair} 
boundary conditions to exist on two opposite edges of the sample. 
It is by now feasible to experimentally
  identify ribbons of graphene with specific boundaries on the atomic
scale. These are preferably of zigzag or of armchair type. For an experimental
realization of our proposal, one would have to look for ribbons 
with semiconducting armchair boundaries.
It is known
that in such a device the valley degeneracy is lifted \cite{brey06,twor06}, 
which is the essential
prerequisite for the appearance of Heisenberg exchange coupling for spins in
tunnel-coupled quantum dots (see below), and thus for the use of graphene 
dots for spin qubits. We show below that spin qubits in graphene can not
only be coupled between nearest neighbor quantum dots but also
over long distances. This long-distance coupling mechanism makes use of 
conduction band to valence band tunnelling processes and is, therefore,
directly based on the Klein paradox in graphene \cite{domb99,kats06}.


%
We now discuss bound-state solutions in our setup, which are required for
a localized qubit. 
We first concentrate on a single quantum dot which is
assumed to be rectangular with width $W$ and length $L$, see 
Fig.~\ref{setup2}. The basic
idea of forming the dot is to take a ribbon of graphene with semiconducting
armchair boundary conditions in $x$-direction and to electrically confine
particles in $y$-direction.
The low energy properties of electrons (with energy $\varepsilon$ with respect
to the Dirac point) in such a setup are described by
the 4x4 Dirac equation
\begin{equation} \label{Dirac}
-i \hbar v \begin{pmatrix}
\sigma_{x}\partial_{x}+\sigma_{y}\partial_{y}&0\\
0&-\sigma_{x}\partial_{x}+\sigma_{y}\partial_{y}
\end{pmatrix}\Psi+eV(y)\Psi=\varepsilon\Psi,
\end{equation}
where the electric gate potential is assumed to vary stepwise,
$V(y) = V_{\rm gate}$ in the dot region (where $0\le y\le L$), and
$V(y) = V_{\rm barrier}$ in the barrier region (where $y<0$ or $y>L$).
In Eq.~(\ref{Dirac}), $\sigma_x$ and $\sigma_y$ are Pauli matrices, $\hbar$ is
Planck's constant divided by $2\pi$, $v$ the Fermi velocity, and $e$ the charge of an electron.
The four component spinor envelope wave function $\Psi =
(\Psi^{(K)}_A,\Psi^{(K)}_B,-\Psi^{(K')}_A,-\Psi^{(K')}_B)$ varies on scales
large compared to the lattice spacing. At this point, we are only
  interested in the
  orbital structure of the wave function. The spin degree of freedom is
  neglected until the final part, 
  where we discuss the Heisenberg exchange
  coupling for spins in tunnel-coupled quantum dots. 
In the wave function $\Psi$, 
$A$ and $B$ refer to the two sublattices in the two-dimensional 
honeycomb lattice of carbon atoms, whereas $K$ and $K'$ refer to the
vectors {\bf K} and {\bf K}' in reciprocal space corresponding to the two
valleys in the bandstructure of graphene.
The appropriate semiconducting
armchair boundary conditions for such a wave function have been formulated in
Ref.~\cite{brey06} and can be written as ($\alpha = A,B$)
\begin{equation} \label{bc}
\Psi_\alpha^{(K)}|_{x=0} =
\Psi_\alpha^{(K')}|_{x=0} , \;\; \Psi_\alpha^{(K)}|_{x=W} = e^{\pm 2\pi/3}
\Psi_\alpha^{(K')}|_{x=W} ,
\end{equation}
corresponding to a width $W$ of the ribbon shown in Fig.~\ref{setup2}, 
where $W$ is not an integer multiple of three unit cells. The $\pm$ signs in
Eq.~(\ref{bc}) (as well as in Eq.~(\ref{qn1}) below) 
correspond to the two possible choices of a number of unit
cells that is not an integer multiple of three.  The full set of plane
wave solutions of Eq.~(\ref{Dirac}) is readily determined \cite{twor06}. It is
well known that the boundary condition (\ref{bc}) yields the following
quantization conditions for the wave vector $k_x \equiv q_n$ in $x$-direction 
\cite{brey06,twor06}
\begin{equation} \label{qn1}
q_n = (n \pm 1/3)\pi/W , \;\; n \in \mathbb{Z} .
\end{equation}
An explicit form of the corresponding wave functions is presented in App.~A
and App.~B. 
The level spacing of the modes (\ref{qn1}) 
can be estimated as $\Delta \varepsilon \approx \hbar v \pi/3W$, which gives 
$\Delta \varepsilon \approx  30 \,{\rm meV}$, where we used that $v \approx 10^6 \, {\rm
  m/s}$ and assumed a quantum dot width of about $W \approx 30 \, {\rm nm}$. 
Note that Eq.~(\ref{qn1}) also determines the energy gap
for excitations as $E_{\rm gap} = 2 \hbar v q_0$. Therefore, this gap is of
the order of 60 meV, which is unusually small for semiconductors. This
is a unique feature of graphene that will allow for long-distance coupling 
of spin qubits as will be discussed below.

We now present in more detail the ground-state solutions, i.e. $n=0$
in Eq.~(\ref{qn1}).
The corresponding ground-state energy $\varepsilon$
can be expressed relative to the
potential barrier $V=V_{\rm barrier}$ in the regions $y<0$ and $y>L$
as $\varepsilon=eV_{\rm barrier} \pm \hbar v (q_{0}^{2}+k^{2})^{1/2}$.
Here, the $\pm$ sign refers to a conduction
band ($+$) and a valence band ($-$) solution to Eq.~(\ref{Dirac}).
For bound states to exist and to decay at $y \rightarrow \pm \infty$, we
require that $\hbar v q_0 > |\varepsilon -eV_{\rm barrier}|$, which implies that 
the wave vector $k_y \equiv k$ in $y$-direction, given by
\begin{equation} \label{kbarrier}
k = i \sqrt{q_0^2 - ((\varepsilon -eV_{\rm barrier})/\hbar v)^2} ,
\end{equation}
is purely imaginary.
In the dot region ($0 \leq y \leq L$), the wave vector $k$ in $y$-direction 
is replaced by $\tilde{k}$, satisfying
$\varepsilon=eV_{\mathrm{gate}}\pm \hbar v (q_0^{2}+\tilde{k}^{2})^{1/2}.$
Again the $\pm$ sign refers to conduction and valence band solutions. 
In the following, we
focus on conduction band solutions to the problem. 

\begin{figure}
\vspace{0.5cm}
\begin{center}
\includegraphics[width=8.5cm]{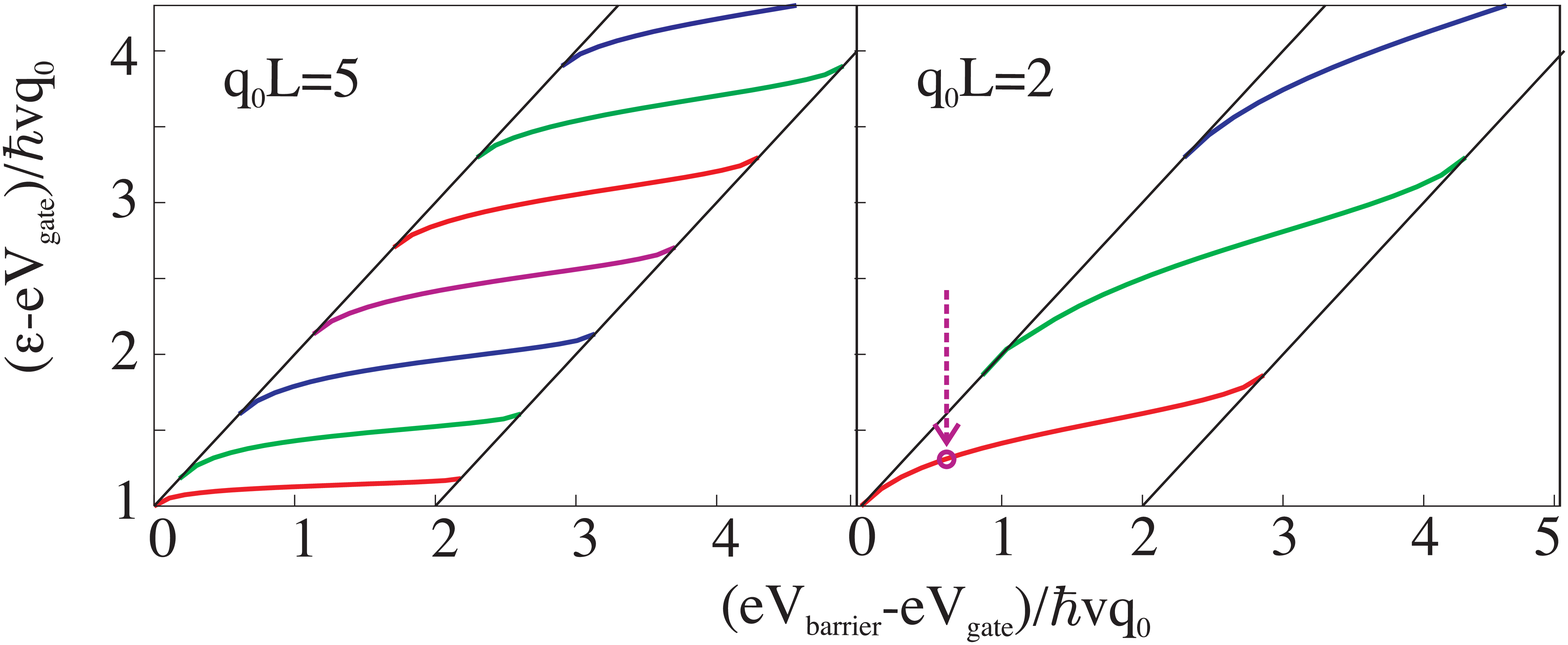} 
\caption{\label{fig2} {\bf Bound state solutions for two different dot sizes.} 
Bound-state solutions of a relatively long ($q_0 L=5$, left panel) 
and a shorter ($q_0 L=2$, right panel) quantum dot are shown. 
The diagonal straight lines mark the area in which 
bound-state solutions can occur. The arrow marks the solution for which the
wave function is plotted in Fig.~\ref{fig4}.}
\end{center}
\end{figure}

\begin{figure}
\vspace{0.5cm}
\begin{center}
\includegraphics[width=8.5cm]{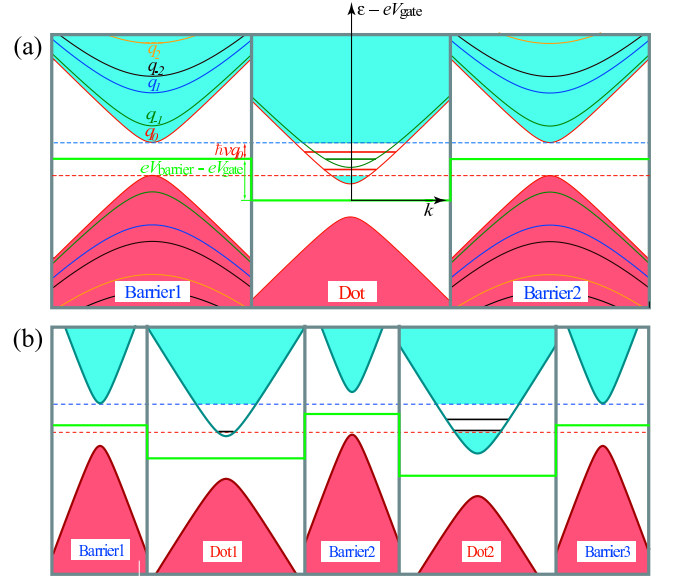} 
\caption{\label{fig3} {\bf Energy bands for single and double dot case.} 
{\bf (a)} Energy bands for two barrier regions and a single dot. 
The red area marks a continuum of states in the
valence bands and the blue area marks a continuum of states in the conduction
bands. In the barrier regions, we indicate the 
energy bands of the quantized
modes due to transverse confinement. All modes 
are non-degenerate solutions in valley
space. They come pairwise in a sense that always two of them are 
separated by a distance $\hbar v q_0$ in energy
space. In the figure, this is illustrated for the energy levels 
corresponding to wave vectors $q_0$ and $q_{-1}$ as well
as $q_1$ and $q_{-2}$. In the dot region, the electric confinement in
longitudinal direction yields an additional level structure, i.e. 
the one shown in
Fig.~\ref{fig2}. For clarity, we only 
show the dot levels that are located in the gap of the barrier
regions and are, therefore, bound states. In the figure, we choose to present
a situation with three bound states in total: Two of them are of 
the $n=0$ series (straight red lines in
the center region) and 
a single one is of the $n=-1$ series 
(straight green line in the center region).
{\bf (b)} Energy bands for a double dot setup. A single bound state (straight
black line) is shown
in the conduction band of the left dot and two bound states are shown in the
conduction band of the right dot. They are coupled via the continuum in the
valence band of the central barrier which is 
enabled by the Klein paradox.}
\end{center}
\end{figure}

Since the Dirac equation (\ref{Dirac}) implies the continuity of the wave
function, the matching condition at $y=0$ and $y=L$ allows us to derive the
transcendental equation for $\varepsilon$
\begin{equation} \label{trans_equation}
e^{2i\tilde{k}L} (z_{0,k} - z_{0,\tilde{k}})^2 - (1- z_{0,k} z_{0,\tilde{k}})^2 = 0
\end{equation}
with $z_{0,k} \equiv (q_0 + i k)/(q_{0}^{2}+k^{2})^{1/2}$.
Eq.~(\ref{trans_equation})
determines the allowed energies $\varepsilon$ for bound states.
In order to analyze the solutions to Eq.~(\ref{trans_equation}), 
we distinguish two cases, one where $\tilde{k}$ is real, and
the other, where $\tilde{k}$ is purely imaginary.  
The two cases are distinguished by the condition
$|\varepsilon -eV_{\rm gate}|\ge \hbar v q_0$ and 
$|\varepsilon -eV_{\rm gate}|< \hbar v q_0$, respectively.
Furthermore, we assume that $V_{\rm gate} \neq V_{\rm barrier}$,
i.e., $z_{0,k}\neq z_{0,\tilde{k}}$. If we relax this assumption, we
  can show that for the case $z_{0,k} = z_{0,\tilde{k}}$ only a single
  solution to Eq.~(\ref{trans_equation}) exists, 
  namely $z_{0,\tilde{k}}=1$,
  which implies that $\tilde{k}=0$. The corresponding wave function to this
  solution vanishes identically (see App.~A for further details). 
In the case where $\tilde{k}$ is
purely imaginary, there is no bound-state solution. This is due to the fact
that such a solution would have to exist directly in the bandgap.
We now analyze solutions for real $\tilde{k}$. In the corresponding 
energy window
\begin{equation} \label{window1} 
|\varepsilon-eV_{\rm gate}| \ge \hbar v q_0 > |\varepsilon-eV_{\rm barrier}| ,
\end{equation}
we can simplify Eq.~(\ref{trans_equation}) considerably, obtaining
\begin{equation} \label{trans1}
\tan(\tilde{k} L) = \frac{\hbar v \tilde{k} 
\sqrt{(\hbar v q_0)^2 - (\varepsilon -eV_{\rm barrier})^2}}{(\varepsilon - eV_{\rm
  barrier})(\varepsilon - eV_{\rm gate}) - (\hbar v q_0)^2} .
\end{equation}
We show a set of solutions to
Eq.~(\ref{trans1}) for a relatively short dot ($q_0 L=2$) as well as a longer dot 
($q_0 L=5$)  in Fig.~\ref{fig2}. The number of bound states $N$ 
(for $n = 0$) is maximal if 
$\Delta V = V_{\rm barrier} - V_{\rm gate}$ is exactly as large as the size of
the gap $E_{\rm gap} = 2 \hbar v q_0$, then
$N_{\rm max} = \left\lceil \sqrt{8} q_0 L/\pi \right\rceil$, 
where $\lceil x\rceil$ is the integer just larger than $x$. The 
level spacing associated with the allowed solutions of Eq.~(\ref{trans1})
increases as $L$ decreases and has a rather complicated parameter
dependence. It can, however, be estimated to be of the order of 
$\Delta \varepsilon \approx \hbar v \pi/{\rm max} \{W,L\}$, which is in the
energy  range of a few tens of meV as mentioned below Eq.~(\ref{qn1}). In
Fig.~\ref{fig3}, we show the energy bands of a single dot and two neighboring
barrier regions as well as a double dot setup with three barrier regions. The
double dot case illustrates how we make use of the 
Klein paradox to couple two dots.

\begin{figure}
\vspace{0.5cm}
\begin{center}
\includegraphics[scale=0.25]{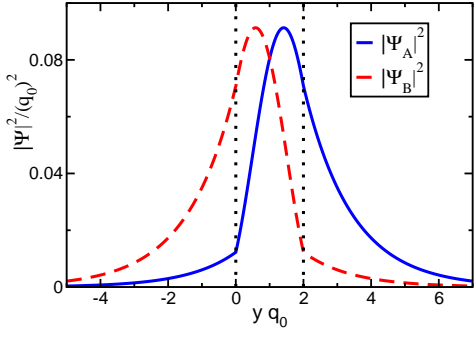} 
\caption{\label{fig4} 
{\bf Ground-state wave function.}
Normalized squared wave functions 
$|\Psi_A|^2 = |\Psi_A^{(K)}|^2 = |\Psi_A^{(K')}|^2$ and 
$|\Psi_B|^2 = |\Psi_B^{(K)}|^2 = |\Psi_B^{(K')}|^2$ for the bound state 
solution for the parameter choice 
$e(V_{\rm barrier}- V_{\rm gate}) = 0.6 \hbar v q_0$, 
and $q_0 L= (\pi/3) L/W = 2$
 (indicated by the arrow in Fig.~\ref{fig2}). 
The corresponding energy is given by 
$\varepsilon \approx 1.31 \cdot \hbar v q_0$. 
The dotted lines indicate the dot region $0 \leq y \leq L$.}
\end{center}
\end{figure}

A particular example of a wave function is shown in Fig.~\ref{fig4}. 
It is 
a ground-state solution under the
parameter choice $e(V_{\rm barrier} - V_{\rm gate}) = 0.6 \hbar v q_0$,
and $q_0 L=2$ (indicated by the arrow in Fig.~\ref{fig2}). The
weight of the wave function on the $A$ and $B$ lattice sites is different,
however, the integrated weight is the same as required by the normalization
condition \cite{brey06}. Ground-state solutions (i.e. the lowest lying (red)
lines in Fig.~\ref{fig2}) have no nodes in the dot region --
similar to the corresponding problem of confined electrons that obey the
non-relativistic Schr{\"o}dinger equation. Excited-state solutions in
parameter regions in which they exist, do have nodes in the dot region, which
is shown in Fig.~8 of App.~C.

\begin{figure}
\vspace{0.5cm}
\begin{center}
\includegraphics[scale=0.375]{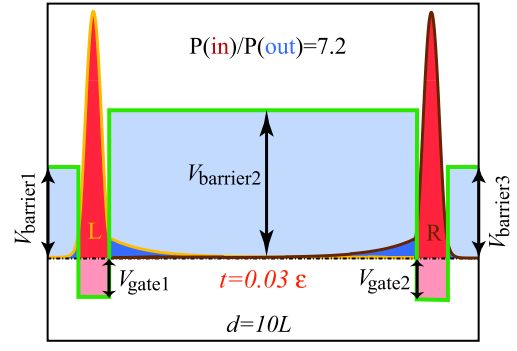} 
\caption{\label{fig5} 
{\bf Long-distance coupling of two qubit ground states.}
The normalized squared wave functions $|\Psi|^2 = 
|\Psi_A^{(K)}|^2 + |\Psi_A^{(K')}|^2 + |\Psi_B^{(K)}|^2 + |\Psi_B^{(K')}|^2$ 
of two qubits separated by a distance
$d=10L$, where $L$ is the length of each quantum dot, are plotted next to each
other. A ground state (of the series with the transverse quantum number $n=0$) 
in the left dot is coupled to a ground state (of the same series with $n=0$) 
in the right dot. The coupling is as large as $t=0.03 \varepsilon$, where
$\varepsilon$ is the ground-state energy. Furthermore, the
qubits are highly localized, which can be seen from the ratio
$P(in)/P(out)$. Here, $P(in)$ is the probability of the electron to be inside
the corresponding dot and $P(out)$ is probability to be outside the dot in the
barrier regions.
The parameters chosen for the potential (in units of $\hbar v q_0/e$) 
are $V_{\rm barrier1}=V_{\rm barrier3}=1$,  $V_{\rm barrier2}=1.65$,
$V_{\rm gate1} = V_{\rm gate2}=-0.5$.}
\end{center}
\end{figure}

We now turn to the case of two coupled graphene quantum dots,
separated by a potential barrier, as sketched in Fig.~\ref{setup2},
each dot filled with a single electron.
It is interesting to ask whether the spins ${\bf S}_i$ of these two electrons
($i=1,2$) are coupled through an exchange coupling, 
$H_{\rm exch}=J{\bf S}_1\cdot{\bf S}_2$,
in the same way as
for regular semiconductor quantum dots \cite{burk99}, because this coupling is,
in combination with single-spin rotations, sufficient to generate
all quantum gates required for universal quantum computation \cite{loss98}.
The exchange coupling is based on the Pauli exclusion principle
which allows for electron hopping between the dots in the spin singlet
state (with opposite spins) of two electrons, but not in a spin triplet
(with parallel spins), thus leading to a singlet-triplet splitting 
(exchange energy) $J$.
However, a singlet-triplet splitting $J\neq 0$ only occurs if the triplet 
state with two electrons on the same dot in the ground state 
is forbidden, i.e., in the case of a
single \textit{non-degenerate} orbital level.
This is a non-trivial requirement in a graphene structure, as in
bulk graphene, there is a two-fold orbital (``valley'') degeneracy
of states around the points ${\bf K}$ and ${\bf K}'$ in the first
Brillouin zone.
This valley degeneracy is lifted in our case of a ribbon with {\it
  semiconducting armchair
edges}, and the ground-state solutions determined by Eq.~(\ref{trans1}) 
are in fact non-degenerate.
The magnitude of the exchange coupling within a Hund-Mulliken
model is \cite{burk99}
$J= (-U_H+ (U_H^2+16t_H^2)^{1/2})/2+V$,
where $t$ is the tunnelling (hopping) matrix element between the 
left and right dot, $U$ is the on-site Coulomb energy, and $V$
is the direct exchange from the long-range (inter-dot) Coulomb interaction.
The symbols $t_H$ and $U_H$ indicate that these quantities are
renormalized from the bare values $t$ and $U$ 
by the inter-dot Coulomb interaction.
For $t \ll U$ and neglecting the long-ranged Coulomb part,
this simplifies to the Hubbard model result
$J=4t^2/U$ where $t$ is the tunnelling (hopping) matrix element between the 
left and right dot and $U$ is the on-site Coulomb energy.
In the regime of weak tunnelling, we can estimate 
$t \approx  \varepsilon  \int \Psi_L^\dagger(x,y)\Psi_R(x,y)  dx \, dy$,
where $\Psi_{L,R}(x,y)=\Psi(x,y \pm (d+L)/2)$ are the ground-state spinor 
wave functions of the left and right dots and $\varepsilon$ is the
single-particle ground state energy.
Note that the overlap integral vanishes if the states on the left and right
dot belong to different transverse quantum numbers $q_{n_L}\neq q_{n_R}$.
For the ground state mode, we have $n_L=n_R=0$, and
the hopping matrix element can be estimated for $d\gtrsim L$ as
\begin{equation} \label{t}
t \approx 4 \varepsilon  \alpha_0 \delta^*_0 W d z_{0,k} \exp(-d |k|) ,
\end{equation}
where $\alpha_0$ and $\delta_0$ are wave function amplitudes (with dimension
1/length) that are specified in App.~C.
As expected, the exchange coupling decreases exponentially with the barrier
thickness, the exponent given by the ``forbidden'' momentum $k$ in
the barrier, defined in Eq.~(\ref{kbarrier}). The amplitude $t$ 
can be maximized by tuning to
a bound-state solution, where $|\varepsilon - eV_{\rm barrier}|$ approaches
$\hbar v q_0$ (from below). Then, $d|k| < 1$. Such 
a fine-tuning can be easily achieved in graphene quantum dots, where the
small band gap allows to sweep through it and, therefore, to 
use conduction and valence band states (of the barrier region) 
to couple quantum dots. In Fig.~\ref{fig3}{\bf (b)}, we sketch the energy
bands for the double dot case which shows how confined states in the
two dots can be coupled via cotunnelling processes through 
the continuum of states in the 
valence band of the central barrier region. Remarkably,
this opens up the possibility for long distance coupling of electron 
spins because, in the limit $|\varepsilon - eV_{\rm
  barrier}| \rightarrow \hbar v q_0$, the coupling $t$ depends only
weakly on the distance $d$ between the quantum dots. However, already for
bound state solutions with $|k|d>1$, a coupling over a length exceeding
several times the dot size is possible. 
For the situation where we couple two ground states in the quantum dots, we 
find, for instance, a
solution, where $|k|d=4$, $d=10 \, L$, and the coupling can still be as
large as $t \approx 0.03 \varepsilon$ for highly localized qubits. This
example is shown in Fig.~\ref{fig5}. If we couple a ground state in the one
dot with an excited state in the other dot, the hopping matrix element 
$t$ can be even larger. The corresponding wave functions for that case 
are illustrated in Fig.~\ref{figd}.
%
The values of $t$, $U$, and $J$ can be estimated as follows.
The tunnelling matrix element $t$ is a fraction of
$\varepsilon\approx 30\,{\rm meV}$ (for a width of $W \approx 30 \,{\rm nm}$), 
we obtain that $t \approx 0.5 \dots 2.5 \,{\rm meV}$.
The value for $U$ depends on screening which we can assume to be
relatively weak in graphene \cite{divi84}, thus, we estimate, e.g.,
$U \approx 10 \,{\rm meV}$, and obtain $J \approx 0.1 \dots 1.5 
\,{\rm meV}$.

%
\begin{figure}
\vspace{0.5cm}
\begin{center}
\includegraphics[scale=0.39]{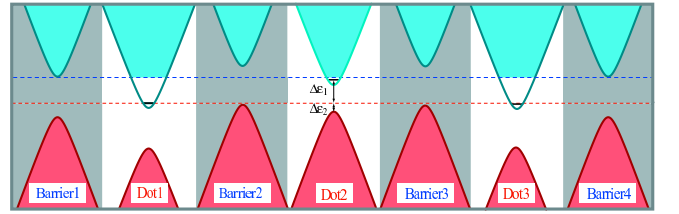} 
\caption{\label{fig6} 
{\bf Triple quantum dot setup.} The energy bands of a triple quantum dot setup
are shown in which dot 1 and dot 3 are strongly coupled via cotunnelling
processes through the valence bands of barrier 2, barrier 3, and dot 2. 
The center dot 2 is decoupled by detuning.  The energy levels are chosen such
that $\Delta \varepsilon_2 \ll \Delta \varepsilon_1$. The triple
dot example illustrates that in a 
line of quantum dots, it is possible to strongly couple any two of them
and decouple the others by detuning. This is a unique feature of graphene and
cannot be achieved in semiconductors such as GaAs that have a much larger gap.}
\end{center}
\end{figure}

For the situation with more than two dots in a line, it turns out that 
we can couple any two of them with the others being decoupled by detuning. 
We mention here that our model is based on a single particle picture. 
Such a model effectively captures effects of Coulomb interactions
as far as they can be described within the constant interaction model 
for quantum dots \cite{grab91}. The Coulomb interaction then only shifts the
energy levels in each dot by a constant. 
In Fig.~\ref{fig6}, we illustrate the situation of three dots in a line where 
the
left and the right dot are strongly coupled and the center dot is decoupled by
detuning. The tunnel coupling of dot 1 and dot 3
is then achieved via Klein tunneling through
the valence band of the two central barriers and the
valence band of the center dot. It is important for the
long-distance coupling that the exchange coupling of
qubit 1 and qubit 3 is primarily achieved via the valence band and not via
the qubit level of the center dot -- leaving the qubit state of dot 2
unchanged. Using the standard transition matrix
approach, we can compare the transition rate of coupling dot 1 and dot 3 via
the continuum of states in the valence band of the center dot (which we call
$\Gamma_{\rm VB}$) with the transition rate via the detuned qubit level of 
the center dot (which we call $\Gamma_{\rm QB}$). We obtain for the ratio 
(see App.~D for the derivation)
\begin{equation} \label{ratio}
\Gamma_{\rm VB}/\Gamma_{\rm QB} \approx (L/W) \ln (4 \Delta/E_{\rm gap}) , 
\end{equation}
where $\Delta \approx 6 \,{\rm eV}$
is the band width of graphene. Therefore, by increasing the aspect ratio 
$L/W$, it is possible to increase the rate $\Gamma_{\rm VB}$ with respect to
$\Gamma_{\rm QB}$. For $L/W=2$ and $E_{\rm gap} = 60 \,{\rm meV}$, 
we find that $\Gamma_{\rm VB}/\Gamma_{\rm QB} \approx 12$, meaning that the
qubit level in dot 2 is barely used to couple
dot 1 and dot 3. This is a unique feature of
graphene quantum dots due to the small and highly symmetric band gap, which is
not known to exist for other semiconducting materials. 
The availability of non-local interactions is important in the context
of quantum error correction, as it raises the error threshold
for fault-tolerant quantum computation \cite{svor05}.
In conclusion, we have proposed a setup to form spin qubits in quantum dots
based on graphene nanoribbons 
with semiconducting armchair boundaries. 
For such a system, we have calculated bound states of a tunable dot and
outlined how two-qubit gates can be realized. We expect
very long coherence times for such spin qubits 
since spin-orbit coupling and hyperfine interaction are known to be weak in
carbon, see App.~E.  Furthermore, we have found that the high flexibility in 
tuning graphene quantum dots in combination with conduction band to valence
band tunnelling based on the Klein paradox 
allows for long distance coupling of electron spins.
Therefore, we propose a system which can serve as the fundamental 
building block for scalable and fault-tolerant quantum computing.

\appendix

\section{General model}

We now present in detail how to derive solutions for bound states in a 
graphene quantum dot. The dot is
assumed to be rectangular with width $W$ and length $L$ as illustrated in
Fig.~1. The basic
idea of forming the dot is to take a strip of graphene with {\it semiconducting
armchair} boundary conditions in $x$-direction and to electrically confine
particles in $y$-direction. Transport properties of a similar system have been
discussed in Ref.~\cite{silv06}. 

The low energy properties of electrons with energy $\varepsilon$ in such a
setup are described by the 4x4 Dirac equation 
\begin{equation} \label{DiracApp}
-i{ \hbar v}  \begin{pmatrix}
\sigma_{x}\partial_{x}+\sigma_{y}\partial_{y}&0\\
0&-\sigma_{x}\partial_{x}+\sigma_{y}\partial_{y}
\end{pmatrix}\Psi+eV(y)\Psi=\varepsilon\Psi,
\end{equation}
with the electric gate potential
\begin{equation}
  \label{potential}
  V(y) = \left\{\begin{array}{l} 
V_{\rm gate},\quad (0\le y\le L),\\
V_{\rm barrier},\quad {\rm otherwise}.
\end{array}
\right.
\end{equation}
In Eq.~(\ref{DiracApp}), $\sigma_x$ and $\sigma_y$ are Pauli matrices $, \hbar$
is Planck's constant devided by $2\pi$, $e$ is the electron charge, and $v$ is
the Fermi velocity. The four component envelope wave function $\Psi =
(\Psi^{(K)}_A,\Psi^{(K)}_B,-\Psi^{(K')}_A,-\Psi^{(K')}_B)$ varies on scales
large compared to the lattice spacing. 
Here, $A$ and $B$ refer to the two sublattices in the two-dimensional 
honeycomb lattice of carbon atoms, whereas $K$ and $K'$ refer to the
vectors in reciprocal space corresponding to the two valleys in the 
bandstructure of graphene.

Plane wave solutions to Eq.~(\ref{Dirac}) take the form \cite{twor06}
\begin{equation}
\Psi^{(+)}_{n,k}(x,y)=\chi^{(+)}_{n,k}(x)e^{iky}, \;\;\; 
\Psi^{(-)}_{n,k}(x,y)=\chi^{(-)}_{n,k}(x)e^{-iky} \label{Psibasisdef}
\end{equation}
with
\begin{eqnarray}
\chi^{(+)}_{n,k}(x)&=&a_{n,+}
\begin{pmatrix}1\\ z_{n,k}\\ 0\\ 0\end{pmatrix}
e^{iq_{n}x}+
a'_{n,+}
\begin{pmatrix}0\\ 0\\ -z_{n,k}\\ 1\end{pmatrix}
e^{iq_{n}x} \nonumber \\
&+& b_{n,+}
\begin{pmatrix}-z_{n,k}\\ 1\\ 0\\ 0\end{pmatrix}
e^{-iq_{n}x}+
b'_{n,+}
\begin{pmatrix}0\\ 0\\ 1\\ z_{n,k}\end{pmatrix}
e^{-iq_{n}x} \nonumber \\
\end{eqnarray}
and
\begin{eqnarray}
\chi^{(-)}_{n,k}(x)&=&a_{n,-}
\begin{pmatrix} z_{n,k}\\ 1 \\ 0\\ 0\end{pmatrix}
e^{iq_{n}x}+
a'_{n,-}
\begin{pmatrix}0\\ 0\\ 1\\ -z_{n,k} \end{pmatrix}
e^{iq_{n}x} \nonumber \\
&+& b_{n,-}
\begin{pmatrix}1\\ -z_{n,k}\\ 0\\ 0\end{pmatrix}
e^{-iq_{n}x}+
b'_{n,-}
\begin{pmatrix}0\\ 0\\ z_{n,k}\\ 1\end{pmatrix}
e^{-iq_{n}x}. \nonumber \\
\end{eqnarray}
The complex number $z_{n,k}$ is given by
\begin{equation} \label{z}
z_{n,k}= \pm \frac{q_n + i k}{\sqrt{k^{2}+q_{n}^{2}}}.
\end{equation}
The energy of the state in the barrier regions ($y<0$ and $y>L$, 
where $V=V_{\rm barrier}$) 
is given by
\begin{equation}
\varepsilon=eV_{\rm barrier} \pm { \hbar v}\sqrt{q_{n}^{2}+k^{2}}.
\end{equation}
In the dot ($0\le y\le L$, where $V=V_{\mathrm{gate}}$) the wave vector $k$ is 
replaced by $\tilde{k}$, satisfying
\begin{equation}
\varepsilon=eV_{\mathrm{gate}}\pm  
{ \hbar v}\sqrt{q_{n}^{2}+\tilde{k}^{2}}.\label{ktildedef}
\end{equation}
The $\pm$ sign in Eqs.~(\ref{z}) -- (\ref{ktildedef}) 
refers to conduction and valence bands. In the following, we
concentrate on conduction band solutions of the problem (keeping in mind that
there is always a particle-hole conjugated partner solution). 

The transverse wave vector $q_{n}$ as well as the coefficients
$a_{n,\pm},a'_{n,\pm},b_{n,\pm},b'_{n,\pm}$ of the $n$-th mode are 
determined (up to a
normalization constant) by the boundary conditions at $x=0$ and $x=W$. We
consider a class of boundary conditions for which the resulting parameters are
independent of the longitudinal wave vectors $k$ and $\tilde{k}$.
We are particularly interested in semiconducting armchair boundary
conditions defined by \cite{brey06} 
\begin{eqnarray}
&&\Psi|_{x=0}=\begin{pmatrix}0&\openone\\ \openone&0\end{pmatrix}
\Psi|_{x=0},\\&&\Psi|_{x=W}=\begin{pmatrix}0&e^{ -i2\pi\mu/3}\openone\\
  e^{ i2\pi\mu/3}\openone&0\end{pmatrix}\Psi|_{x=W},  \label{armchairbc2}
\end{eqnarray}
where $\mu=\pm1$ is defined by the width of the graphene strip 
$W=a_0(3M+\mu)$, with $M$ a positive integer  
($a_0=0.246\:$nm is the graphene lattice constant),
and $\openone$ is the 2x2 unit matrix.
A strip whose 
width is an integer multiple of three unit cells ($\mu=0$) is 
metallic and not suitable for spin qubit applications. The states of a 
semiconductor strip are non-degenerate (in valley space):
\begin{equation} \label{seq1}
q_{n}=\frac{\pi}{W}(n+\mu/3) , \;\; n \in \mathbb{Z}
\end{equation} 
with
$a_{n,\pm}=b'_{n,\pm}=0$, $a'_{n,\pm}=b_{n,\pm}$ (for $\mu=-1$) or
$a'_{n,\pm}=b_{n,\pm}=0$, $a_{n,\pm}=b'_{n,\pm}$ (for $\mu=1$).
 Note that $q_n$ 
determines the size of the gap for each
mode $n$ that is due to the boundary conditions. The size of the gap of mode
$n$ is given by $2{  \hbar v}q_n$. 
For  concreteness, we consider the case of $\mu=1$ only. It can be shown that for the case of $\mu=-1$, the bound states and normalized squared wave functions have exactly the same dependence on the parameters of a quantum dot, $V_\mathrm{barrier}$, $V_\mathrm{gate}$, $W$, and $L$.

Our ansatz for a bound state solution at energy $\varepsilon$ 
to Eq.~(\ref{Dirac}) then reads
\begin{equation}
\Psi= \left\{\begin{array}{ll}
\alpha'_{n} \chi^{(-)}_{n,k}(x)e^{-iky}, &\mathrm{if}\;\;y<0,\\
\beta'_{n} \chi^{(+)}_{n,\tilde{k}}(x)e^{i\tilde{k}y}+
\gamma'_{n} \chi^{(-)}_{n,\tilde{k}}(x)e^{-i\tilde{k}y}, &\mathrm{if}\;\;0\le y\le L,\\
\delta'_{n} \chi^{(+)}_{n,k}(x)e^{ik(y-L)}, &\mathrm{if}\;\;y>L.
\end{array}\right. \label{wf_general_bis}\\
\end{equation}
For bound states, the wave function should decay for $y \rightarrow \pm
\infty$, so we require that
\begin{equation} \label{k}
k = i \sqrt{q_n^2 - (\varepsilon - { e}V_{\rm barrier})^2{ /(\hbar v)^2}} , 
\end{equation}
where ${ \hbar v}q_n > |\varepsilon - { e}V_{\rm barrier}|$ always has to hold.

To find bound state solutions, we have to analyze the 
following set of equations
(coming from wave function matching at $y=0$ and $y=L$)
\begin{eqnarray}
\alpha_n \begin{pmatrix} z_{n,k} \\ 1 \end{pmatrix} &=& \beta_n 
\begin{pmatrix} 1 \\ z_{n,\tilde{k}} \end{pmatrix} + \gamma_n 
\begin{pmatrix} z_{n,\tilde{k}} \\ 1 \end{pmatrix} , \label{eqs1} \\
\delta_n \begin{pmatrix} 1 \\ z_{n,k} \end{pmatrix} &=& \beta_n 
\begin{pmatrix} 1 \\ z_{n,\tilde{k}} \end{pmatrix} e^S + \gamma_n 
\begin{pmatrix} z_{n,\tilde{k}} \\ 1 \end{pmatrix} e^{-S} \nonumber ,
\end{eqnarray}
where $\alpha_n = \alpha'_n a_{n,-}$, $\beta_n = \beta'_n a_{n,+}$, 
$\gamma_n = \gamma'_n a_{n,-}$, $\delta_n = \delta'_n a_{n,+}$, and
$S \equiv i \tilde{k} L$. We can write Eq.~(\ref{eqs1}) as
\begin{equation}
\begin{pmatrix} z_{n,k} & -1 & -z_{n,\tilde{k}} & 0 \\
1 & -z_{n,\tilde{k}} &  -1 & 0 \\
0 & -e^S & - z_{n,\tilde{k}} e^{-S} & 1 \\
0 & -z_{n,\tilde{k}} e^{S} & - e^{-S} & z_{n,k}
\end{pmatrix}
\begin{pmatrix} \alpha_n \\ \beta_n \\ \gamma_n \\ \delta_n \end{pmatrix} 
= 0 .
\end{equation}
The allowed energy values $\varepsilon$ 
are readily determined by finding the roots of the determinant of the matrix
on the lhs of the latter equation
\begin{equation} \label{the_equation}
e^{2S} (z_{n,k} - z_{n,\tilde{k}})^2 - (1- z_{n,k} z_{n,\tilde{k}})^2 = 0 .
\end{equation}

A rather obvious solution is
$z_{n,\tilde{k}}=1$ (which implies that $S=0$) corresponding to
\begin{equation} \label{obsol}
\varepsilon = \pm  { \hbar v}q_n + eV_{\rm gate} .
\end{equation}
However, the corresponding wave functions to the allowed energy
solutions (\ref{obsol}) vanish identically. So, in order to proceed, we
have to find other (less trivial) solutions to the transcendental equation 
(\ref{the_equation}). Since the case $z_{n,k}=z_{n,\tilde{k}}$ only has the 
trivial solution (\ref{obsol}), we can assume that $z_{n,k} \neq
z_{n,\tilde{k}}$, which means that $V_{\rm barrier} \neq V_{\rm gate}$.
Then, we find that
\begin{equation}
\label{eig_equation}
e^{S} = \pm \frac{1- z_{n,k} z_{n,\tilde{k}}}{z_{n,k} - z_{n,\tilde{k}}} .
\end{equation}
To analyze the solutions of Eq.~(\ref{eig_equation}), we distinguish two cases,
one where
\begin{equation} \label{S}
S=i\tilde{k}L
\end{equation}
is purely imaginary (i.e., $\tilde{k}$ is real), and
another, where $S$ is real.  The two cases are distinguished by the criterion
$|\varepsilon -eV_{\rm gate}| \ge  { \hbar v}q_n$ and 
$|\varepsilon -eV_{\rm gate}|< { \hbar v} q_n$, respectively.
In the former case, since the lhs of Eq.~(\ref{eig_equation}) has modulus 
unity, also the rhs must be unimodular, which is satisfied if in addition 
$|\varepsilon-{ e}V_{\rm barrier}| \le { \hbar v} q_n$.  This case, where the equation for the argument of Eq.~(\ref{eig_equation}) remains to be solved,
is discussed in Sec.~\ref{Simag}.  The latter case where $S$ is real
has no solutions. 
 
Indeed, let us rewrite Eq.~(\ref{eig_equation}) as follows:
\begin{equation}
e^{2S}-1=\frac{-2\tilde{k}k}{q_n^2-(\varepsilon-eV_\mathrm{barrier})(\varepsilon-eV_\mathrm{gate})/(\hbar v)^2+\tilde{k}k}.
\label{eq:kappa}
\end{equation}
Taking into account that $q_n^2>(\varepsilon-eV_\mathrm{barrier})(\varepsilon-eV_\mathrm{gate})/(\hbar v)^2$ and $\tilde{k}k\in\mathbb{R}$, we find that the left and right sides of this equation have different signs, therefore,  Eq.~(\ref{eig_equation}) has no roots for any purely imaginary $\tilde{k}$.

\section{Bound state solutions}
\label{Simag}

We now restrict ourselves to the energy window
\begin{equation} \label{energywindow}
|\varepsilon-eV_{\rm gate}| \ge { \hbar v} q_n \ge |\varepsilon-eV_{\rm barrier}| .
\end{equation} 
Then $\tilde{k}$
is real, therefore $|e^{S}|=1$ and $|z_{n,\tilde{k}}|=1$. Furthermore,
$z_{n,k}$ is real. We define $z_{n,\tilde{k}} \equiv e^{i \theta_n}$, 
where
\begin{equation}
\theta_n = \arctan(\tilde{k}/q_n) .
\end{equation}
It is easy to verify that in the energy window (\ref{energywindow})
\begin{equation}
\left|\frac{1- z_{n,k} z_{n,\tilde{k}}}{z_{n,k} - z_{n,\tilde{k}}}\right|=1 .
\end{equation}
We can now rewrite Eq.~(\ref{eig_equation}) as
\begin{equation} \label{trans1app}
\tan(\tilde{k} L) = \frac{\sin \theta_n (1-z_{n,k}^2)}{2 z_{n,k} - 
(1+z_{n,k}^2)\cos \theta_n}
\end{equation}
and further simplify this expression by using that
\begin{eqnarray}
\sin \theta_n &=& \frac{\tilde{k}/q_n}{\sqrt{1+(\tilde{k}/q_n)^2}} , \\
\cos \theta_n &=& \frac{1}{\sqrt{1+(\tilde{k}/q_n)^2}} .
\end{eqnarray}
After some algebra, we obtain
\begin{equation} \label{transeq}
\tan(\tilde{k} L) = \frac{-i \tilde{k} k}{(\varepsilon - { e}V_{\rm barrier})
(\varepsilon - { e}V_{\rm gate}){ /(\hbar v)^2}  - q_n^2} .
\end{equation}
The latter equation in combination with Eq.~(\ref{k}) yields Eq.~(7). 
Numerical solutions to Eq.~(\ref{transeq}) are shown in Fig.~2.

By applying different voltages to the gate and the barriers we shift the 
energy bands of the graphene ribbon under the barriers with respect to 
that of the quantum dot. A bound state in the quantum dot is allowed once 
the energy of the state hits the band gap of the barriers. 
If the difference of the barrier and gate voltages 
$\Delta V=|V_\mathrm{barrier}-V_\mathrm{gate}|$ is less than the energy of 
the gap $2{ \hbar v} |q_n|$ for $n$-th subband, $\tilde{k}$ of a bound 
state lies in the interval $[-\tilde{k}_\mathrm{max},\tilde{k}_\mathrm{max}]$, 
where $\tilde{k}_\mathrm{max}$ is found from the condition 
$\varepsilon(\tilde{k}_\mathrm{max})=V_\mathrm{barrier}+{ \hbar v} |q_n|$ and, 
therefore, 
$\tilde{k}_\mathrm{max}=e\Delta V\sqrt{1+2{ \hbar v} |q_n|/e\Delta V}$. 
The number of bound states (for a given subband index $n$) 
is proportional to the length of the quantum 
dot $L$ and is given by 
$N=\lceil \tilde{k}_\mathrm{max} L/\pi \rceil$ ($\lceil x\rceil$ is the 
integer just larger than $x$) . The number of the bound states of the 
$n$-th subband is maximal, when the barrier-gate voltage difference 
equals the energy band gap ($\Delta V=2{ \hbar v} |q_n|$), and so
$N_\mathrm{max}=\lceil \sqrt{8}|q_n|L/\pi\rceil$. In the case of 
$\Delta V>2{ \hbar v} |q_n|$, the top of the valence band of the 
graphene ribbon under the barriers becomes higher than the bottom of 
the conduction band of the quantum dot, therefore, there are no bound 
states with energies $\varepsilon<eV_\mathrm{barrier}-{ \hbar v} |q_n|$ 
and $|\tilde{k}|$ of a bound state lies in the interval 
$[\tilde{k}_\mathrm{min},\tilde{k}_\mathrm{max}]$, where 
$\tilde{k}_\mathrm{min}$ is found from the condition 
$\varepsilon(\tilde{k}_\mathrm{min})=V_\mathrm{barrier}-{ \hbar v} 
|q_n|$ ($\tilde{k}_\mathrm{min}=e\Delta V\sqrt{1-2{ \hbar v} 
|q_n|/e\Delta V}$), therefore, bound states lie in the energy window 
$eV_\mathrm{barrier}-{ \hbar v} |q_n|\le\varepsilon\le 
eV_\mathrm{barrier}+{ \hbar v} |q_n|$ (as shown in Fig.~2) 
and the number of the bound states is given by 
$N=\lceil \tilde{k}_\mathrm{max} L/\pi \rceil-\lceil \tilde{k}_\mathrm{min} L/\pi \rceil$.

With increasing the barrier-gate voltage difference, a $m$-th bound state
appears at $\Delta V_0=-{ \hbar v} |q_n|+{ \hbar v}
\sqrt{q_n^2+(\pi/L)^2(m-1)^2}$ with the energy
$\varepsilon_m^{(0)}=V_\mathrm{gate}+\Delta V_0+{ \hbar v} |q_n|$ and ends up
at $\Delta V_1={ \hbar v} |q_n|+{ \hbar v} \sqrt{q_n^2+(\pi/L)^2m^2}$ with the
energy $\varepsilon_m^{(1)}=V_\mathrm{gate}+\Delta V_1-{ \hbar v} |q_n|$ (see
Fig.~2). 


\section{Wave function}
\label{sec_wf}

\begin{figure}
\vspace{1.0cm}
\begin{center}
\includegraphics[scale=0.25]{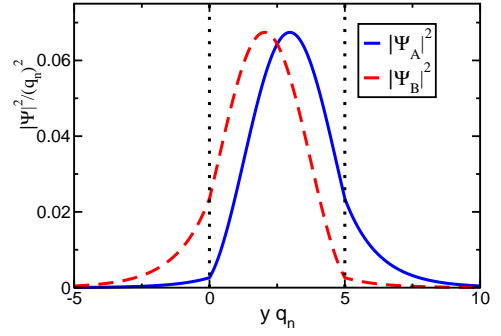} 
\caption{\label{fig1_sm} {\bf Ground-state wave function.} 
Normalized squared wave function
$|\Psi_A|^2=|\Psi^{(K)}_A|^2= |\Psi^{(K')}_A|^2$ and 
$|\Psi_B|^2=|\Psi^{(K)}_B|^2= |\Psi^{(K')}_B|^2$ for the ground 
state solution of a dot of length $q_n L=5$ 
with corresponding energy $\varepsilon=1.101\:{ \hbar v}   q_n$. 
Here, $e(V_{\rm barrier} - V_{\rm gate}) = 0.5\:{ \hbar v} q_n$.
The dotted lines indicate the dot region
$0 \leq y q_n \leq 5$.}
\end{center}
\end{figure}

\begin{figure}
\vspace{1.0cm}
\begin{center}
\includegraphics[scale=0.25]{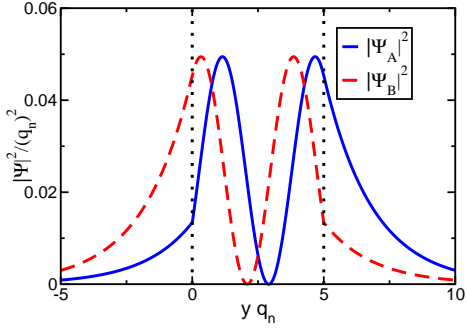} 
\caption{\label{fig2_sm} {\bf Excited-state wave function.}
Normalized squared wave function 
$|\Psi_A|^2=|\Psi^{(K)}_A|^2= |\Psi^{(K')}_A|^2$ and 
$|\Psi_B|^2=|\Psi^{(K)}_B|^2= |\Psi^{(K')}_B|^2$
for the 
first excited 
state solution of a dot of length $q_n L=5$ 
with corresponding energy $\varepsilon=1.34\: { \hbar v} q_n$. 
Here, $e(V_{\rm barrier} - V_{\rm gate}) = 0.5\:{ \hbar v} q_n$.
The dotted lines indicate the dot region
$0 \leq y q_n \leq 5$.}
\end{center}
\end{figure}

Following Brey and Fertig \cite{brey06}, we write the wave function as 
\begin{equation} \label{Psi1}
\Psi (x,y) = \begin{pmatrix} \Psi_A^{(K)} (x,y) \\ \Psi_B^{(K)}
  (x,y) \\ - \Psi_A^{(K')} (x,y) \\ - \Psi_B^{(K')} (x,y) 
\end{pmatrix}
\end{equation}
and give solutions for each component separately. As mentioned above, 
the subscripts $A$ and $B$
refer to the two sublattices in the two-dimensional honeycomb lattice of
carbon atoms and the superscripts $K$ and $K'$ refer to the two valleys in
graphene. Note that the normalization condition \cite{brey06}
\begin{equation} \label{norma}
\int d dx\, dy\, \Bigl[ | \Psi_\mu^{(K)} (x,y)|^2 + 
| \Psi_\mu^{(K')} (x,y)|^2 \Bigr]=\frac{1}{2}
\end{equation}
(for $\mu=A,B$ each) finally determines the normalization constant of the wave
function. With the ansatz (\ref{wf_general_bis})
we explicitly obtain the following components of the wave function
\begin{eqnarray}
\Psi_A^{(K)} (x,y) &=& \left\{\begin{array}{l}
\alpha_n z_{n,k} e^{i q_n x} e^{-i k y} , \\
\beta_n e^{i q_n x} e^{i \tilde{k} y} + \gamma_n z_{n,\tilde{k}} e^{i q_n x} 
e^{-i \tilde{k} y} , \\
\delta_n e^{i q_n x} e^{i k (y-L)} ,
\end{array}\right. \label{psiAK} \\
\Psi_B^{(K)} (x,y) &=& \left\{\begin{array}{l}
\alpha_n e^{i q_n x} e^{-i k y} , \\
\beta_n z_{n,\tilde{k}} e^{i q_n x} e^{i \tilde{k} y} + \gamma_n e^{i q_n x} 
e^{-i \tilde{k} y} , \\
\delta_n z_{n,k} e^{i q_n x} e^{i k (y-L)} ,
\end{array}\right. \label{psiBK} \\
- \Psi_A^{(K')} (x,y) &=& \left\{\begin{array}{l}
\alpha_n z_{n,k} e^{-i q_n x} e^{-i k y} , \\
\beta_n e^{-i q_n x} e^{i \tilde{k} y} + \gamma_n z_{n,\tilde{k}} e^{-i q_n x} 
e^{-i \tilde{k} y} , \\
\delta_n e^{-i q_n x} e^{i k (y-L)} ,
\end{array}\right. \label{psiAKp} \\
- \Psi_B^{(K')} (x,y) &=& \left\{\begin{array}{l}
\alpha_n e^{-i q_n x} e^{-i k y} , \\
\beta_n z_{n,\tilde{k}} e^{-i q_n x} e^{i \tilde{k} y} + \gamma_n e^{-i q_n x} 
e^{-i \tilde{k} y} , \\
\delta_n z_{n,k} e^{-i q_n x} e^{i k (y-L)} .
\end{array}\right. \label{psiBKp}
\end{eqnarray}
In the latter equations, the first line corresponds to the region in space,
where $y<0$, the second line to $0\le y\le L$, and the third line to $y>L$.
Thus, we obtain that
\begin{eqnarray}
|\Psi_A^{(K)} (x,y)|^2 &=& |\Psi_A^{(K')} (x,y)|^2 , \\
|\Psi_B^{(K)} (x,y)|^2 &=& |\Psi_B^{(K')} (x,y)|^2 .
\end{eqnarray}
We now plot the normalized squared wave function of a ground-state solution
and a excited-state solution of a dot with length $q_n L =5$ in
Figs.~\ref{fig1_sm} and \ref{fig2_sm}, respectively. These are obtained from
Fig.~2 under the choice that $e(V_{\rm barrier} - V_{\rm
  gate}) = 0.5 \hbar v q_n$. 
Evidently, the ground-state solution has no nodes in the
dot region, whereas the excited-state solution has nodes.

\section{Long-distance coupling}

\begin{figure}
\vspace{1.0cm}
\begin{center}
\includegraphics[scale=0.4]{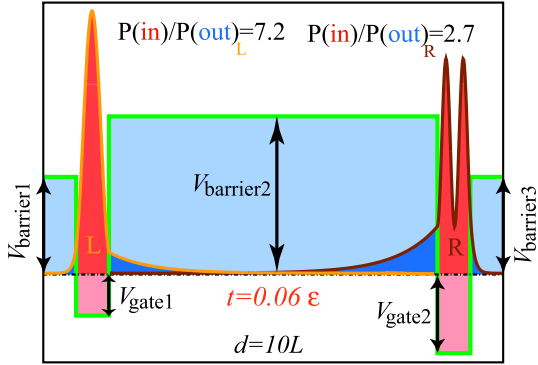} 
\caption{\label{figd} {\bf Long-distance coupling of a ground state and an
    excited state.}
The normalized squared wave functions $|\Psi|^2 = 
|\Psi_A^{(K)}|^2 + |\Psi_A^{(K')}|^2 + |\Psi_B^{(K)}|^2 + |\Psi_B^{(K')}|^2$ 
of two qubits separated by a distance
$d=10L$, where $L$ is the length of each quantum dot, are plotted next to each
other. A ground state (of the series with the transverse quantum number $n=0$) 
in the left dot is coupled to an excited state (of the same series with $n=0$) 
in the right dot. The coupling is as large as $t=0.06 \varepsilon$, where
$\varepsilon$ is the ground-state energy of the left dot. Furthermore, the
qubits are still highly localized, which can be seen from the ratio
$P(in)/P(out)$. Here, $P(in)$ is the probability of the electron to be inside
the corresponding dot and $P(out)$ is probability to be outside the dot in the
barrier regions.
The parameters chosen for the potential (in units of $\hbar v q_0/e$) 
are $V_{\rm barrier1}=V_{\rm barrier3}=1$,  $V_{\rm barrier2}=1.65$,
$V_{\rm gate1}=-0.5$, and $V_{\rm gate2}=-0.9$.}
\end{center}
\end{figure}

\subsection{Long-distance coupling of two qubits} 

Here, we discuss a particular example of long-distance coupling of two qubits
separated by a distance $d$.
The coupling is achieved via a continuum of states in the valence band of the
barrier region as shown in Fig.3{\bf(b)} of the Letter. Therefore, the
long-range coupling is enabled by the Klein paradox. In the weak tunneling
regime, the hopping matrix element is given by
\begin{equation} \label{tapp}
t \approx  \varepsilon  \int \Psi_L^\dagger(x,y)\Psi_R(x,y)  dx \, dy ,
\end{equation}
where $\Psi_{L,R}(x,y)=\Psi(x,y\pm (d+L)/2)$ are the spinor 
wave functions of the left and right dots and $\varepsilon$ is the
single-particle energy of the coupled levels. The integration in transverse
$x$-direction is trivial and just gives a factor $W$. The
integration in longitudinal $y$-direction can be restricted to the integration
window $y \in [-d/2,d/2]$ 
if the wave functions are predominantly localized in the dot
regions. Then, the hopping matrix element can be estimated for $d\gtrsim L$ as
\begin{equation} \label{tbis}
t \approx 4 \varepsilon  \alpha_0 \delta^*_0 W d z_{0,k} \exp(-d |k|) ,
\end{equation}
where $\alpha_0$ and $\delta_0$ are wave function amplitudes specified in
Eqs.~(\ref{psiAK}) -- (\ref{psiBKp}). In Eq.~(\ref{tbis}), we assumed that
only levels of the series corresponding to the $n=0$ transverse mode are
coupled. It is easy to relax this assumption because, if higher
transverse modes form bound states, then only modes with $n_L=n_R$
contribute to $t$, where $n_{L/R}$ is the transverse quantum number in the
left/right dot. In Fig.~\ref{figd}, we demonstrate that a rather large
coupling of $t=0.06 \varepsilon$ can be achieved over a distance as large as
ten times the size of the quantum dots (see also Fig.~5 for
comparison). Note that the qubits in this example
are well localized in the corresponding dot regions: The probability of the
electron in the left dot to be in the dot region $P(in)$ is 7.2 times larger
than to be in the barrier regions $P(out)$. For the right dot, the ratio of
$P(in)/P(out)=2.7$ is a bit smaller but the electron is still predominantly
localized in the dot region. 

\subsection{Long-distance coupling in multiple quantum dot setup }

In Fig.6, we propose a triple quantum dot setup in which dot 1
and dot 3 are strongly coupled and the center dot 2 is decoupled by
detuning. It is important that dot 1 and dot 3 are coupled via the valence
band states of dot 2 and not via the (detuned) qubit level of dot
2. Otherwise, the spin of the decoupled qubit level would be affected by the
coupling of the other qubits which is unwanted in the proposed long-distance
coupling scheme. 
We assume that the gates that put the three dots in the Coulomb
blockade regime are set in such a way that cotunnelling processes from dot 1
via dot 2 to dot 3 happen in the following order: First, an electron tunnels
from dot 2 to dot 3 and then an electron tunnels from dot 1 to dot 2. The
system is described by the Hamiltonian
\begin{equation}
H = H_0 + H_T ,
\end{equation}
where the kinetic term describes three qubit levels ($\alpha = 1,2,3$) 
and the continuum of states in the valence band of dot 2
\begin{equation} \label{hnull}
H_0 = \sum_{\alpha = 1-3} \sum_{\sigma = \uparrow, \downarrow}
E_{\alpha,\sigma} a^\dagger_{\alpha,\sigma} a_{\alpha,\sigma} +
\sum_{k} \sum_{\sigma= \uparrow, \downarrow}
\varepsilon_{k,\sigma} b^\dagger_{k,\sigma} b_{k,\sigma}
\end{equation}
and the tunnelling Hamiltonian reads
\begin{eqnarray} \label{ht}
H_T &=& t \sum_\sigma \left(a^\dagger_{1,\sigma} a_{2,\sigma} + 
a^\dagger_{2,\sigma} a_{3,\sigma} + \;\; {\rm H.c.}\right) \nonumber \\
&+& t \sum_{k,\sigma}\left( a^\dagger_{1,\sigma} b_{k,\sigma} + 
b^\dagger_{k,\sigma} a_{3,\sigma} + \;\; {\rm H.c.} \right) .
\end{eqnarray}
In Eq.~(\ref{hnull}), $a_{\alpha,\sigma}$ and $b_{k,\sigma}$ annihilate
electrons with spin $\sigma$ in the qubit level of dot $\alpha$ and in the
valence band of dot 2, respectively. We assume that $E_{1,\sigma} =
E_{3,\sigma}$ (i.e. qubit 1 and qubit 3 are on resonance) and
$\Delta \varepsilon_1 = E_{2,\sigma} - E_{1,\sigma} \approx E_{\rm gap} = 2
\hbar v q_0$ (see Fig.~6). 
In Eq.~(\ref{ht}), we make the approximation that all tunnelling
matrix elements $t$ depend only very weakly on energy and are real. 
The transmission rate
from an initial state $|i \rangle$ to a final state $|f \rangle$ can be
calculated using Fermi's golden rule
\begin{equation}
W_{fi} = \frac{2\pi}{\hbar} |\langle f | T(\varepsilon_i) |i \rangle |^2
\delta (\varepsilon_f - \varepsilon_i)
\end{equation}
with the transition matrix given by (up to second order in $H_T$)
\begin{equation} \label{tmatrix}
T(\varepsilon) = H_T + H_T \frac{1}{\varepsilon + i \eta -H_0} H_T + \dots .
\end{equation}
We can put $\eta=0$ in the latter equation because we are only interested in
off-resonant cotunnelling processes. This means that $\Delta \varepsilon_2$ 
(see Fig.~6) should be finite because we want to have
well-localized qubit states. The corresponding matrix elements of the T-matrix
(\ref{tmatrix}) may be written as
\begin{equation}
(T(\varepsilon))_{k,k'} = (H_T)_{k,k'} + \sum_{k''} (H_T)_{k,k''}
\frac{1}{\varepsilon - \varepsilon_{k''}} (H_T)_{k'',k'} + \dots
\end{equation}
Now, we want to calculate $|T_{13}|^2 \equiv 
|\langle 3 | T(E_{1,\sigma}) | 1 \rangle|^2$, where
$| 1 \rangle$ and $| 3 \rangle$ are the ground states of qubit $1$ and $3$,
respectively. The lowest non-vanishing contribution of $|T_{13}|^2$ is of
cotunnelling type, i.e., of fourth order in $t$. 
It is possible to separate the different contributions to 
$|T_{13}|^2$ into three terms, namely
\begin{equation} \label{t13}
|T_{13}|^2 = |T^{(QB)}_{13}|^2 + |T^{(VB)}_{13}|^2 + 2 \, 
{\rm Re} \, [(T^{(QB)}_{13})^* T^{(VB)}_{13}] .
\end{equation}
In the latter equation, $|T^{(QB)}_{13}|^2$ determines the transition rate via
the qubit level of dot 2 (the unwanted process), $|T^{(VB)}_{13}|^2$
determines the transition rate via the continuum of states in the 
valence band of dot 2 (the wanted
process), and $2 \, {\rm Re} \, [(T^{(QB)}_{13})^* T^{(VB)}_{13}]$ is the
interference term of the two paths.
It is straightforward to derive that (for a given spin $\sigma$ of qubit 2)
\begin{equation}
T^{(QB)}_{13} (E_{1,\sigma}) = \frac{t^2}{E_{1,\sigma} - E_{2,\sigma}} .
\end{equation}
This has to be compared with
\begin{eqnarray} \label{t13vb}
T^{(VB)}_{13} (E_{1,\sigma}) 
&=& t^2 \int_{E_{\rm gap}/2}^\Delta dE
\frac{\nu_0(E)}{E_{1,\sigma} - E} \nonumber \\
&\approx& - \frac{L t^2}{\pi \hbar v} \ln (4 \Delta/E_{\rm gap}) ,
\end{eqnarray}
where
\begin{equation}
\nu_0(E) = \frac{L}{\hbar v \pi} \frac{E}{\sqrt{E^2-(E_{\rm gap}/2)^2}}
\end{equation}
is the density of states of the mode $n=0$ with $E_{\rm gap}=2\hbar v q_0$. 
In Eq.~(\ref{t13vb}), we integrate over the whole band width of the valence
band (bounded by $\Delta \approx 6 \,{\rm eV}$). The approximate result in the second
line of Eq.~(\ref{t13vb}) holds for the hierarchy of energies 
$\Delta \gg E_{\rm gap} \gg E_{1,\sigma}$. (In a more general case, the
integral in Eq.~(\ref{t13vb}) can still be evaluated analytically but yields a
less compact expression.)  

The contribution coming from $|T^{(QB)}_{13}|^2$ is evidently the smallest
term of the three terms on the rhs of Eq.~(\ref{t13}). If we want to compare
the rate that does not affect the spin of the qubit level in dot 2 (which we
call $\Gamma_{\rm VB}$) with the largest of the rates that does affect the spin
of the qubit level in dot 2 (which we call $\Gamma_{\rm QB}$), we can estimate
\begin{equation}
\frac{\Gamma_{\rm VB}}{\Gamma_{\rm QB}} = \frac{|T^{(VB)}_{13}|^2}{|2 \, {\rm
    Re} \, [(T^{(QB)}_{13})^* T^{(VB)}_{13}]|} \approx \frac{L}{W} \ln (4
\Delta/E_{\rm gap}) .
\end{equation}
The latter result is Eq.~(9). It shows that by increasing the
aspect ratio $L/W$ we can increase the weight of the coupling via the valence
band states of dot 2 (which is wanted) as compared to the weight of the
coupling via the qubit state of dot 2 (which is unwanted).

\section{Decoherence}

Finally, we give some arguments and rough estimates for the 
spin decoherence in graphene.
It is generally believed that spin-orbit effects are weaker in carbon 
than in GaAs due to the lower atomic weight.  Therefore, the dominating
mechanism for decoherence will be the hyperfine coupling to the nuclear
spins that are present in the material.

The coherence time given by the hyperfine coupling can be
estimated as \cite{cois04} $t \approx \sqrt{N/c} A^{-1}$, 
where $N$ is the number of atoms in the dot
(for a typical graphene dot, $N\approx 10^4$),
$c N$ is the number of atoms per dot with a nuclear spin,
and $A$ the coupling constant of the hyperfine interaction. 
Since we mainly deal with $\pi$ orbitals in graphene,
the contact hyperfine interaction is strongly reduced
and the hyperfine interaction is dominated by its dipolar part.
From the best available calculations for the dipolar hyperfine matrix
elements \cite{tang00,antr93}, one obtains
\begin{equation}
A_{\rm dip} =  \frac{8}{5}\frac{\mu_0}{4\pi} \mu_B \mu 
\left\langle\frac{1}{r^3}\right\rangle
\approx 0.38 \,\mu{\rm eV},
\end{equation}
where $\mu$ is the nuclear magneton of $^{13}{\rm C}$,
$\mu_B$ is the Bohr magneton, and $\mu_0$ the
vacuum dielectric constant.
This estimated value for  $A_{\rm dip}$ is 
smaller than $A_{\rm GaAs}\approx 90\,\mu {\rm eV}$
by more than two orders of magnitude.
The natural abundance of $^{13}{\rm C}$ is about $c\approx 1\%$ which
yields, with the values quoted above, a coherence time of approximately
$t\approx 10\,\mu{\rm s}$, about a thousand times longer than in GaAs.
Unlike in GaAs, this value can be improved by isotopic 
purification.  Reducing the $^{13}{\rm C}$ content by
a factor of about 100 already decreases the average number of
nuclear spins per dot to about one.  This allows for
a preselection of the dots without any nuclear spin to
be used as qubits.

{\bf Acknowledgements} 
We thank H.A. Fertig and L.M.K. Vandersypen 
for discussions and acknowledge support from
the Swiss NSF, NCCR Nanoscience, DARPA, ONR, and JST ICORP.

\end{document}